\documentclass[preprints,article,accept,moreauthors,pdftex,10pt,a4paper]{mdpi}
\usepackage{amssymb}
\usepackage[normalem]{ulem}

\firstpage{1} 
\pubvolume{xx}
\issuenum{1}
\articlenumber{1}
\pubyear{2020}
\copyrightyear{2020}
\externaleditor{}
\history{}

\Title{Optimal interplay between synaptic strengths and network structure enhances activity fluctuations and information propagation in hierarchical modular networks}

\Author{Rodrigo F.O. Pena $^{1}$\orcidA{}, Vinicius Lima $^{1}$\orcidB{}, Renan O. Shimoura$^{1}$\orcidC{}, Jo\~ao P. Novato$^{1}$, and Antonio C. Roque$^{1,*}$\orcidD{}}

\AuthorNames{Rodrigo Pena, Vinicius Lima, Renan Shimoura, Jo\~ao Novato, and Antonio Roque}

\address{%
$^{1}$ \quad Dept of Physics, Faculty of Philosophy, Sciences and Letters of Ribeir\~{a}o Preto, University of S\~{a}o Paulo, Ribeir\~{a}o Preto, SP, Brazil \\
}

\corres{Correspondence: antonior@usp.br}

\abstract{In network models of spiking neurons, the joint impact of network structure and synaptic parameters on activity propagation is still an open problem. Here we use an information-theoretical approach to investigate activity propagation in spiking networks with hierarchical modular topology. We observe that optimized pairwise information propagation emerges due to the increase of either (i) the global synaptic strength parameter or (ii) the number of modules in the network, while the network size remains constant. At the population level, information propagation of activity among adjacent modules is enhanced as the number of modules increases until a maximum value is reached and then decreases, showing that there is an optimal interplay between synaptic strength and modularity for population information flow. This is in contrast to information propagation evaluated among pairs of neurons, which attains maximum value at the maximum values of these two parameter ranges. By examining the network behavior under increase of synaptic strength and number of modules we find that these increases are associated with two different effects: (i) increase of autocorrelations among individual neurons, and (ii) increase of cross-correlations among pairs of neurons. The second effect is associated with better information propagation in the network. Our results suggest roles that link topological features and synaptic strength levels to the transmission of information in cortical networks.}

\keyword{hierarchical modular networks; cortical network models; neural information processing; delayed transfer entropy; neural activity fluctuations}

\begin{document}


\section{Introduction}

Neurons in the cerebral cortex are interconnected according to selective, i.e. non-random, patterns of connectivity. Different experimental procedures are advancing the knowledge on these intricate connectivity patterns (see e.g. \cite{Paxinos1999,sporns2005,Bullmore2011,sporns2011,alivisatos2013,daCosta2013,Stephan2013,Szalkai2019}). With the help of computational models, the improved connectivity maps are allowing the realization of the long-standing goal of understanding the interplay between structure and dynamics in cortical networks \cite{potjans2014,schuecker2017,yamamoto2018}. Yet, it is an open question whether the evolutionary process which generated such a complex cortical wiring is the result of a selection mechanism for optimized region-to-region communication or some higher-order function \cite{Laughlin2003,Tkacik2016,Avena2018}.

Connectivity may follow different classification schemes beyond physical (structural) connectivity per se. Functional and effective connectivity, which respectively relate to statistical dependencies among neural activity in different brain regions and causal influence of one brain region over another are widely used but captured by different procedures \cite{friston2011, van2010}. Independently of the connectivity scheme used, experimental studies generally agree that cortical networks have hierarchical modular architecture \cite{mountcastle1997,hagmann2008,BulSpo09,kaiser2010,meunier2010,shafi2018}. Previous works have shown that this type of architecture allows long-lived self-sustained activity states in spiking network models with characteristics akin to cortical spontaneous activity patterns \cite{wang2011,TomPen14,TomPen16}. However, these studies have not addressed the effect of the hierarchical modular architecture on information flow in the network.

Other studies based on network models with non hierarchical modular architectures have investigated the information processing capability of the network by playing with other features. Examples are the strength of the global synaptic coupling parameter in random networks with sparse connectivity \cite{ostojic2014}; the degree of synchronization among pools of excitatory and inhibitory neurons connected by feedback loops \cite{Buehlmann2010}; and, in the context of reservoir computing \cite{lukosevicius2009}, the community structure within the reservoir \cite{rodriguez2019}, and the presence of topographically structured feed-forward connections within the reservoir \cite{Zajzon2019}.   

The question of how topology is connected to information transmission is appealing specially due to recent anatomical developments \cite{shih2015}, where it was shown that pathways of information flow in the \emph{Drosophila} connectome can be predicted from the network structure, or more theoretically oriented ones \cite{rodriguez2019}, where the authors showed that an intermediate level of modularity in artificial recurrent neural networks is optimal for memory performance. Indeed, there is a general agreement that architecture shapes communication \cite{Zajzon2019}.

In this work, we tackle the problem of information transmission in hierarchical modular networks of spiking neurons. We study networks of different levels of hierarchical organization, which determines the number of modules, and overall strength of synaptic coupling. Using information-theoretical measures we show that information transmission in these networks have different dependencies on the level of hierarchy and the synaptic coupling strength. By analyzing information transmission between neurons and between modules we show that the latter is not straightforwardly predictable from the former, disclosing the complexity behind communication dynamics in hierarchical modular networks. In particular, we find that there is an intermediate range of number of modules (neither too few nor too many) for which information transmission between modules is maximal. This ``optimality" phenomenon is not observed for information transmission between neurons. Our results underscore the importance of the hierarchical modular architecture of the cortex and suggest an interplay between network structure and synaptic strength with consequences for cortical information transmission.

\section{Methods}

\subsection{Neuron Model}
We use the leaky integrate-and-fire (LIF) neuron model \cite{gerstner2014}:

\begin{equation}
\label{Eq:LIF}
\tau_{\rm m} \dot{v}_j = -v_j + R\left(I_{j, \rm loc} + I_{j, \rm ext}\right),
\end{equation}

\noindent where $v_j$ is the membrane potential of neuron $j$, $R$ is the membrane resistance and $\tau_{\rm m}$ is the membrane time constant in ms. The synaptic currents arriving at neuron $j$ are represented by $I_{j, \rm loc}$, which represents the ``local'' input, and $I_{j, \rm ext}$, which represents the external input received by neuron $j$. This model obeys a fire-and-reset rule so that when the voltage reaches the threshold $v_{\rm th}$ a spike is considered to be emitted and the voltage is reset to the reset potential $v_{\rm r}$. We also consider a refractory period of duration $\tau_{\rm ref}$ after a spike for which the neuron is unable to respond. 

Upon arrival of an excitatory input to neuron $j$, $RI_{j, \rm loc}$ is incremented by $J$ (in mV) and upon arrival of an inhibitory input it is incremented by $-gJ$, where $g$ is the relative inhibitory synaptic strength parameter. Synaptic communication has a delay of $\tau_{\rm D}$, which is the same for all neuron pairs. The single neuron and network parameters are shown in Table~\ref{Tab::parameters}.

\begin{table}[!htp]
	\begin{center}
		\begin{tabular}{ccc}
			\multicolumn{3}{c}{\textbf{PARAMETERS}}\tabularnewline
			\hline 
			\multicolumn{3}{c}{\rule{0pt}{2.6ex} \textbf{Neuron parameters}}\tabularnewline
			\hline 
			\rule{0pt}{2.1 ex}\textbf{Name} & \textbf{Value} & \textbf{Description}\tabularnewline
			$\tau_{\rm m}$ & 20 ms & Membrane time constant\tabularnewline
			$v_{\rm th}$ & 20 mV & Firing threshold\tabularnewline
			$v_{\rm r}$ & 10 mV & Reset potential\tabularnewline
			$\tau_{\rm R}$ & 0.5 ms & Refractory period\tabularnewline
			$RI_{\rm ext}$ & 30 mV & External input\tabularnewline
			\hline 
			\multicolumn{3}{c}{\rule{0pt}{2.6ex} \textbf{Network connectivity parameters}}\tabularnewline
			\hline 
			\rule{0pt}{2.1 ex}\textbf{Name} & \textbf{Value} & \textbf{Description}\tabularnewline
			$N$ & $ 2^{17} $ & Size of excitatory population\tabularnewline
			$\epsilon$ & $0.01$ & Connectivity\tabularnewline
			$R_{\rm ex}$ & $0.9$ & Excitatory rewiring probability\tabularnewline
			$R_{\rm in}$ & $1$ & Inhibitory rewiring probability\tabularnewline
			\hline 
			\multicolumn{3}{c}{\rule{0pt}{2.6ex} \textbf{Synaptic parameters}}\tabularnewline
			\hline 
			\rule{0pt}{2.1 ex}\textbf{Name} & \textbf{Value} & \textbf{Description}\tabularnewline
			$J$ & $\in [0;1]$ mV & Excitatory synaptic strength\tabularnewline
			$g$ & 5 & Relative inhibitory synaptic strength\tabularnewline
			$\tau_{\rm D}$ & $0.55$ ms & Synaptic delay\tabularnewline
			\hline 
			
		\end{tabular}
		\caption{\label{Tab::parameters}Summary of parameters used in this paper.
		}
	\end{center}
\end{table}

\subsection{Network}

The hierarchical modular networks used here are constructed as described below \cite{wang2011,TomPen14,TomPen16}. We start with a random network of $N=2^{17}=131,072$ neurons connected with connectivity $\epsilon=0.01$. The parameter $\epsilon$ is the probability of a synaptic connection between any pair of neurons in the network. The ratio of excitatory to inhibitory neurons is 4:1. This network has only one module and will be called a network of hierarchical level $H$=0. Networks of higher hierarchical levels are generated by the following algorithm:
\begin{enumerate} \item Randomly divide each module of the network into two modules of equal size;  \item With probability $R_{\rm ex/in}$, replace each {\em intermodular} connection $i\to j$ by a new connection between $i$ and $k$ where $k$ is a randomly chosen neuron from {\em the same module} as $i$;  
\item Recursively apply steps 1 and 2 to build networks of higher ($H$=2,3$\ldots$) hierarchical levels. A network with hierarchical level $H$ has $2^H$ modules.  
\end{enumerate} 
The rebating probabilities have values $R_{\rm ex}= 0.9$ and $R_{\rm in}= 1$, so that the intermodular connections are exclusively excitatory.  

\begin{figure*}[!htp]
	\centering
	\includegraphics[scale=.3]{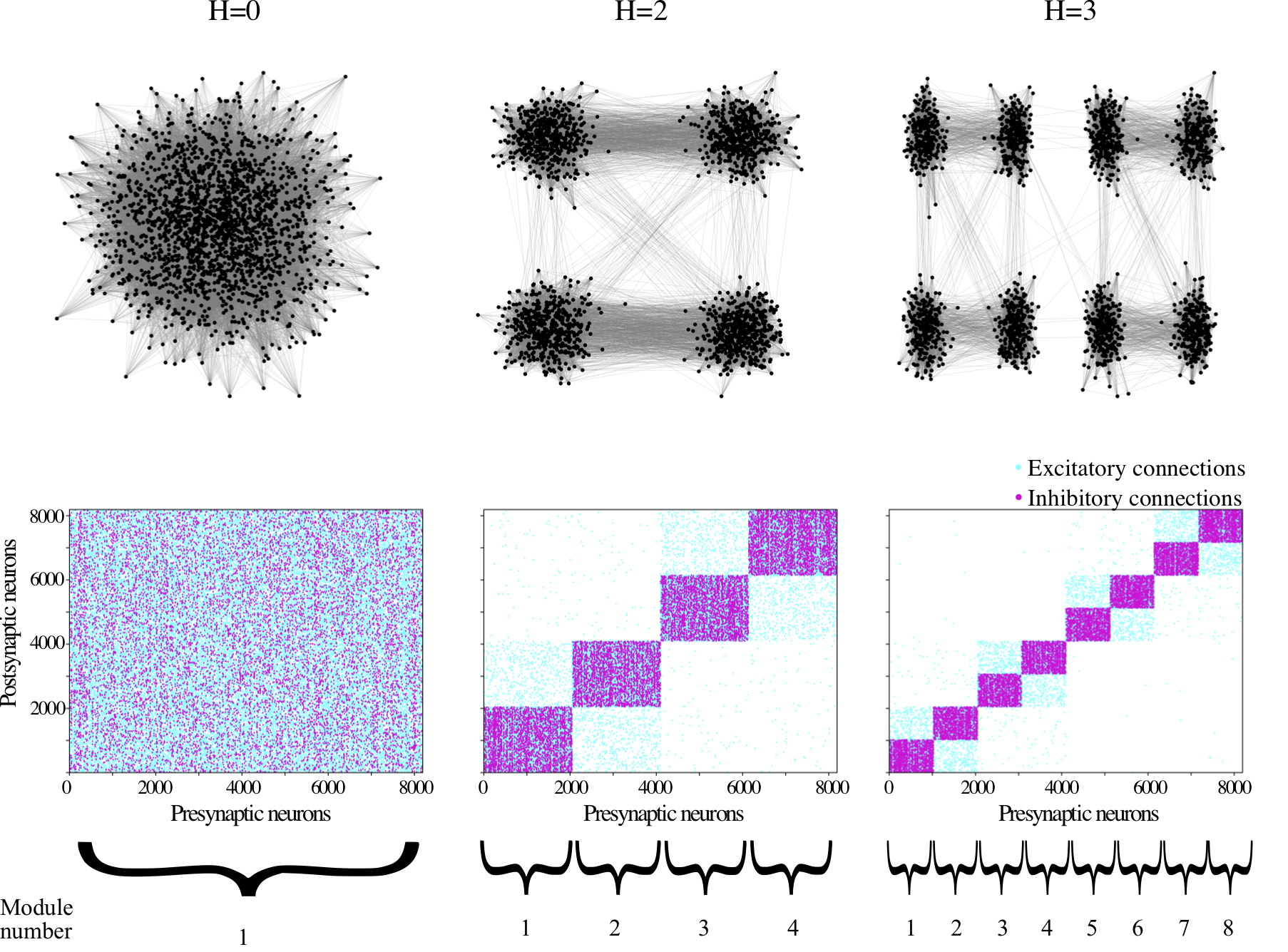}
	\caption{\textbf{Examples of hierarchical modular networks of different hierarchical levels}. Upper row: Schematic representation of the network for $H =$ 0, 2 and 3. In the figures, only networks with $N = 2^{11}$ and exclusively excitatory neurons were used for ease of visualization and to highlight the intermodular connections. Bottom row: Adjacency matrices for networks with $N = 2^{13}$ neurons (excitatory and inhibitory in the 4:1 ratio) and the same $H$ levels as in the top row. Each dot represents a connection from a presynaptic neuron to a postsynaptic neuron. Blue dots represent excitatory neurons and red dots represent inhibitory neurons. For each hierarchical level $H$, the module numbers are shown below the corresponding adjacency matrix.}
	\label{Fig:HMN}
\end{figure*}

Some examples of hierarchical modular networks are shown in Fig.~\ref{Fig:HMN}. They allow a visualization of the hierarchical structure of the network: as $H$ increases, the number of modules increase and modules are encapsulated in groups of modules. Connections between modules that are ``topologically" closer are denser than between more topologically distant ones. Inhibitory connections occur strictly within modules (are ``local") while excitatory connections can be both local and long-range. For purposes that will be described below, we introduce an arbitrary ordering scheme for modules (see the bottom of Fig.~\ref{Fig:HMN}).  

\subsection{Simulation protocol}

We study hierarchical modular networks with hierarchical level $H$ in the range [0,9], where $H=0$ corresponds to a network with Erd\H{o}s-R{\'e}nyi topology (see above). For each $H$ level, the network is submitted to the same stimulation protocol, aimed at simulating spontaneous activity in the network. The stimulation protocol consists of applying a constant external input $RI_{\rm ext} = 30$ mV to all neurons of the network for the simulation time $T=2$ sec. 

For each $H$ level, the above stimulation protocol was repeated for coupling strengths $J$ in the range [0,1] with increments of 0.05. The value of $g$ was fixed at 5 for all simulations. The network activity in each simulation was characterized by the statistical measures described below.

\subsection{Statistics}

The spike train of neuron $j$ is given by the sum of delta functions:

\begin{equation}
\label{Eq:spike_train}
x_j(t) = \sum_i \delta(t - t_i^f),
\end{equation}

\noindent where $t_i^f$ is the time of the $i$th spike of neuron $j$. From the spike train, one can obtain the firing rate of neuron $j$ over a time interval $T$ as $\nu_ j=\langle x_j(t) \rangle = n_j/T = \left( \int_{T} x_{j}(t) dt \right)/T$.

The network time-dependent firing rate (activity) of a population of $N$ neurons is defined as
\begin{equation}
r(t; \Delta t) = \frac{1}{N \Delta t} \sum_{j=1}^{N} \int_{t}^{t+\Delta t} x_j(t') dt',
\label{Eq:firing_rate}
\end{equation}
\noindent where the time window is fixed at $\Delta t = 1$ ms. For simplicity, below we will denote this time-dependent firing rate by $r(t)$. The average of $r(t)$ over a time interval $T$ will be indicated here by $\nu$.

The power spectrum of $x_j(t)$ is defined as:
\begin{equation}
\label{Eq:power}
S_{xx,j}(f) = \frac{\langle \tilde{x}_j(f)\tilde{x}_j^*(f)\rangle}{T},
\end{equation}

\noindent where $T$ is the simulation time and $\tilde{x}_j(f)$ is the Fourier transform of the $j$th spike-train given by $\tilde{x}_j(f)=\int_0^T dt e^{2\pi i f t} x_j(t)$ and $\tilde{x}_j^*(f)$ is its complex conjugate.

In general, we consider the averaged spike-train power spectrum over a number $K$ of neurons
\begin{equation}
\label{Eq:power_avg}
\bar{S}_{xx}(f) = \frac{1}{K}\sum_{j \in K}S_{xx,j}(f).
\end{equation}

To evaluate the spike train’s long-term variability we use the Fano factor ($FF$),
\begin{equation}
\label{Eq:fano_factor}
FF= \langle \Delta n^2 \rangle / \langle n \rangle,
\end{equation}

\noindent where $n$ is the spike count defined as $n=\int_0^T x(t) dt$ for a given time window $T$. A large value of $FF$ indicates an enhancement of slow fluctuations. In our simulations, we extract $FF$ from $\bar{S}_{xx}(f)$ since both are related by the equation:  $\lim\limits _{f\to 0}\bar{S}_{xx}(f)= \nu \times FF$. From $\bar{S}_{xx}(f)$ we also extract the mean firing-rate of the network by the relationship: $\lim\limits _{f\to \infty}\bar{S}_{xx}(f)= \nu$ (cf. \cite{grun2010,pena2018}). 

For spike-trains we compute the autocorrelation function  
\begin{equation}
\label{Eq:autocorr}
c_{xx}(\tau) =  \frac{1}{K}\sum_{j \in K} \left( \langle x_j(t)x_j(t+\tau) \rangle - \langle x_j(t) \rangle \langle x_j(t+\tau) \rangle \right),
\end{equation}

\noindent which in our work is always an average over $K=10,000$ randomly chosen neurons and normalized by $c_{xx}(0)$. Similarly, the cross-correlation function $c_{xy}(\tau)$ is computed by taking $K=10,000$ randomly chosen pairs of spike-trains $x(t)$ and $y(t)$.

Following \cite{NeiYak07,wieland2015,pena2018}, we also extract the correlation time $\tau_{c}$ from $\bar{S}_{xx}(f)$ by means of the Parseval theorem applied to the integral over the squared and normalized correlation function 
\begin{equation}
\label{Eq:tau_c}
\tau_c = \int_{-\infty}^{+\infty}d\tau \left[\frac{\hat{c}(\tau)}{\hat{c}(0)} \right]^2 = \int_{-\infty}^{+\infty} df \frac{(\bar{S}_{xx}(f) - \nu)^2}{\nu^4},
\end{equation}

\noindent where $\hat{c}(\tau)$ denotes the continuous part of the spike train’s correlation function,
\begin{equation}
\label{Eq:norm_corr}
\hat{c}(\tau) = \underbrace{\left( \langle x(t)x(t+\tau) \rangle - \langle x(t) \rangle \langle x(t+\tau) \rangle \right)}_{\text{correlation function } c(\tau)} - \nu \delta(\tau).
\end{equation}

To measure information flow in the network we make use of the Transfer Entropy ($TE$) \cite{Sch00}. This quantity measures how much the predictability of the spike train $x(t)$ of a given neuron is improved if we have knowledge about the spike train $y(t)$ of a different neuron \cite{palmigiano2017} (for simplicity we will denote the spike-trains at a given time $t$ by $x_t$ and $y_t$). 

Given that the measure is asymmetric it also conveys a directional sense, i.e. whether information is flowing from $x$ to $y$ or vice-versa.

Here we use a version of $TE$ called delayed transfer entropy \cite{hansen2011}, which is given by 

\begin{align}
TE_{y \rightarrow x}(d) = \sum p(x_{t+1+d}, x_{t+d}, y_{t}) \log_{2}\left(\frac{p(x_{t+1+d}, x_{t+d}, y_{t})p(y_{t})}{p(y_{t+1},y_{t})p(x_{t},y_{t})}\right).
\label{Eq:delayed_transfer_entropy}
\end{align}

\noindent Equation~\ref{Eq:delayed_transfer_entropy} refers to the situation when a presynaptic neuron $y$ sends signals to a postsynaptic neuron $x$. In this case, $TE_{y \rightarrow x}(d)$ is obtained by taking four spike-trains: $y_{t}$, $x_{t}$, the spike train of the receiving neuron shifted by a delay $d$ ($x_{t+d}$), and the spike train of the receiving neuron shifted by delay $d+1$ ($x_{t+d+1}$). From these spike-trains, we determine the probability $p(y_{t})$, the joint probabilities $p(y_{t+1},y_{t})$, $p(x_{t},y_{t})$, and $p(x_{t+1+d}, x_{t+d}, y_{t})$, which are used to calculate $TE_{y \rightarrow x}(d)$. In Eq.~\ref{Eq:delayed_transfer_entropy}, the summation is taken over the set of all possible combinations of symbols for the spike-trains. 

\begin{figure}[!htp]
	\centering
	\includegraphics[width=1\textwidth]{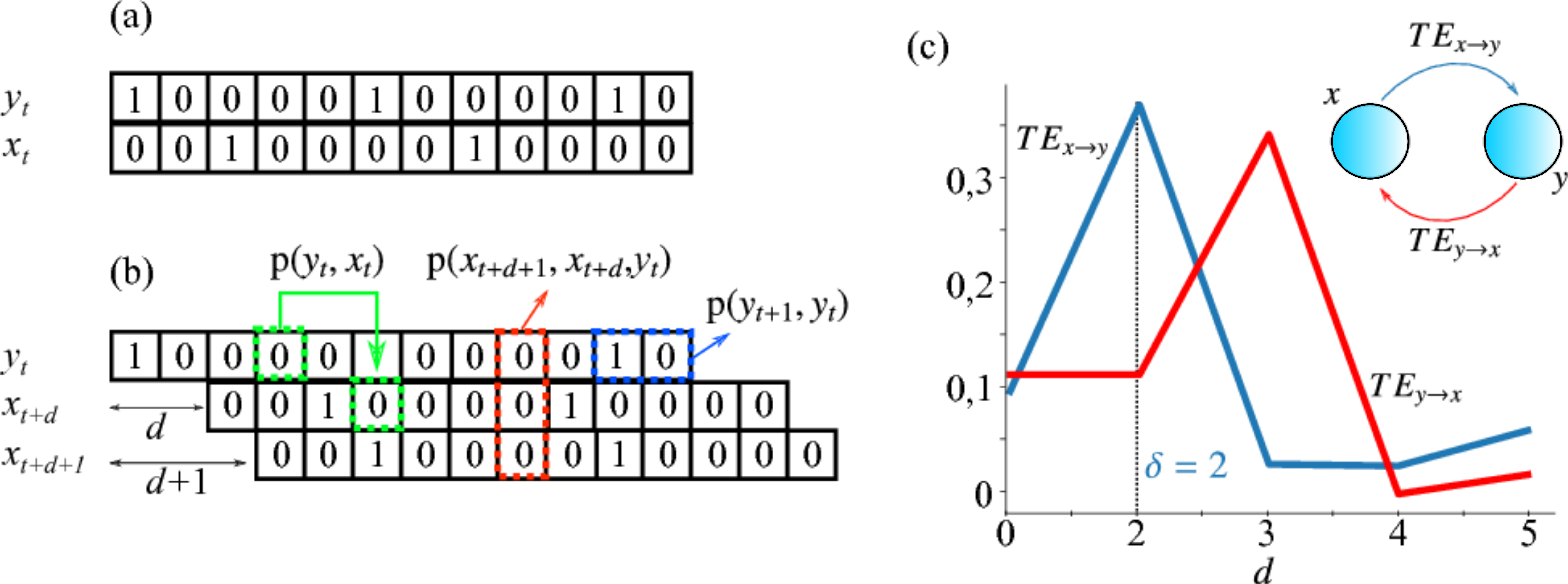}
	\caption{\textbf{Method to measure the delayed transfer entropy using the joint probability distributions.} (a) First we take two spike trains of a pair of neurons in the network. (b) Then we apply a delay $d$ in one of them to determine the joint probability distributions $p(x_{t},y_{t})$ (indicated by the green arrow), $p(x_{t+1+d}, x_{t+d}, y_{t})$ (indicated by the red arrow), and $p(y_{t+1},y_{t})$ (indicated by the blue arrow). Next we estimate the transfer entropy by inserting these distributions into Eq.~\ref{Eq:delayed_transfer_entropy}. (c) Example plots of $TE_{y \rightarrow x}$ and $TE_{x \rightarrow y}$ for a simple system of two coupled neurons (shown in the inset) with $x \rightarrow y$ connection delay $\delta_{x \rightarrow y}=2$ and $y \rightarrow x$ connection delay $\delta_{y \rightarrow x}=3$. The respective $TE$s are maximized when the measure delay $d$ is the same as the corresponding connection delay.}
	\label{Fig::te_measure_method}
\end{figure}

Since the value of the spike-train in each time step is either $0$ (for silence) or $1$ (for a spike), for the joint probabilities $p(x_{t},y_{t})$ we have $2^2 = 4$ combinations, and for $p(x_{t+1+d}, x_{t+d}, y_{t})$ we have $2^3 = 8$ combinations. In Fig.~\ref{Fig::te_measure_method} we summarize the procedure to measure $TE_{y \rightarrow x}$ explained above. In Fig.~\ref{Fig::te_measure_method}(a) the spike-trains were made in such a way that whereas $TE_{y \rightarrow x}$ is maximum for $d=3$, $TE_{x \rightarrow y}$ is maximum for $d=2$. To illustrate that $TE$ is maximized when the delay is equal to the time delay of the connection between two neurons and that this measure is asymmetric ($TE_{y \rightarrow x} \neq TE_{x \rightarrow y}$), in Fig.~\ref{Fig::te_measure_method}(c) we plot $TE_{y \rightarrow x}$ and $TE_{x \rightarrow y}$ for a simple network of two coupled neurons. The system was artificially set up so that $x$ fires three time steps after $y$ and $y$ fires two time steps after $x$. The delay for which $TE$ is maximum can be interpreted not only as the time that information takes to go from $y$ to $x$ but also as the time delay of a possible functional connection between the pair of neurons \cite{wibral2015}. In fact, many studies use this approach to determine and retrieve the connectivity map of a network \cite{deabril2018}.

For each combination of the parameters $\{J,H\}$ we compute the network $TE$ by selecting $K=10,000$ randomly chosen combinations of neuron pairs (neuron $y$ and neuron $x$) without repetition. For each pair, $TE$ is measured as in Eq.~\ref{Eq:delayed_transfer_entropy}; since the communication delay is unknown we measure $TE$ for delays in the range $d \in [155; 300]$ bins, with bin size of $0.1$ ms, and use the maximum $TE$ in this range  \cite{wibral12013}. The choice of range for bins was made taking into consideration the synaptic delay time $\tau_D$ and the membrane time constant $\tau_{\rm m}$ (which characterizes the voltage rise time towards spike threshold). In the end, we extract the average $TE$, 
\begin{equation}
\langle TE \rangle = \frac{1}{K} \sum_{j \in K} \max \{ TE_j(d) \},
\label{Eq:TE_mean}
\end{equation}

\noindent where $TE_{j}$ is the transfer entropy for the $j$th pair of neurons. Considering that we used $100$ different combinations of $\{J,H\}$ for $10$ different initial conditions (yielding $1000$ networks), and that we used $10,000$ neuron pairs over a range of $145$ delays, there were at least $1.45$ billion computations to obtain $\langle TE \rangle$ in this work. Thus, the computation of $\langle TE \rangle$ demanded extensive parallel computation.

The above definition of $TE$ is valid for spike trains of neurons pairs. It will be called here "microscopic" $TE$, or simply $TE$. We introduce here a second definition of $TE$, based on firing rates (activities) of pairs of modules, which will be used to measure information flow at the macroscopic level. We will refer to this "macroscopic" $TE$ as $TE^{(H)}$. To calculate $\langle TE^{(H)} \rangle$ for a given hierarchical level $H$, we randomly select 500 pairs of modules and measure the transfer entropy for each pair $(i,j)$ using equation~\ref{Eq:delayed_transfer_entropy} with $d=0$ and $x$ and $y$ being the activities $r_{i}(t)$ and $r_{j}(t)$ of the two modules, respectively. The activity of a module is calculated as in equation~\ref{Eq:firing_rate} with $N$ equal to the number of neurons in the module. Then, we take the average over the 500 pairs of modules,
\begin{equation}
\langle TE^{(H)} \rangle = \frac{1}{K} \sum_{j=1}^{K} TE_j^{(H)}
\label{Eq:TE_macro}
\end{equation}
\noindent where $j$ is the index of the module pair, $TE_j^{(H)}$ is the transfer entropy for the $j$th pair, and $K=500$. For networks with less than $500$ combinations of modules we compute $\langle TE^{(H)} \rangle$ as above but taking the average over the smaller number of module pairs. Since the activity of a module is continuous we estimated the joint probabilities in equation~\ref{Eq:delayed_transfer_entropy} using a Gaussian kernel density estimator with bandwidth 0.3 \cite{Sch00}.

To evaluate statistical dependency among modules, we extract the mutual information  \cite{deabril2018} among pairs of adjacent modules using a procedure similar to the one described above for $\langle TE^{(H)} \rangle$. The mutual information between two variables $x$ and $y$ is given by:
	
\begin{equation}
MI(x;y) = \sum_{\substack{x \in x_t\\  y \in y_t}} p(x,y) \log_{2} \frac{p(x,y)}{p(x)p(y)}.
\end{equation}	

\noindent For a given hierarchical level, we select the $2^{H}$ pairs of adjacent modules $\{ (1,2), (2,3), \ldots, (2^{H}-1, 2^{H}), (2^{H}, 1) \}$, where the numbering scheme is the one introduced in Fig.~\ref{Fig:HMN}. Then, the mean mutual information over the set of $2^H$ adjacent modules is given by $\langle MI^{(H)} \rangle = \sum^{2^H}_{k=1}MI_k/2^H$, where $MI_k$ is the mutual information between the $k$th pair of adjacent modules as defined above.

All neuron and network models were implemented using the Brian 2 neurosimulator \cite{stimberg2019brian}. Statistical and information theoretical analyses were implemented by self-developed Python packages which are made available at GitHub \cite{codes}. Network visualization was made with the help of the Python package NetworkX.  Simulations were performed with the use of the NeuroMat (\url{neuromat.numec.prp.usp.br/}) cluster. 

\section{Results}

\subsection{Information transfer is enhanced when both modularity and synaptic strength increase}\label{sec:results_sec1}

\begin{figure*}[!htp]
	\centering
	\includegraphics[scale=0.22]{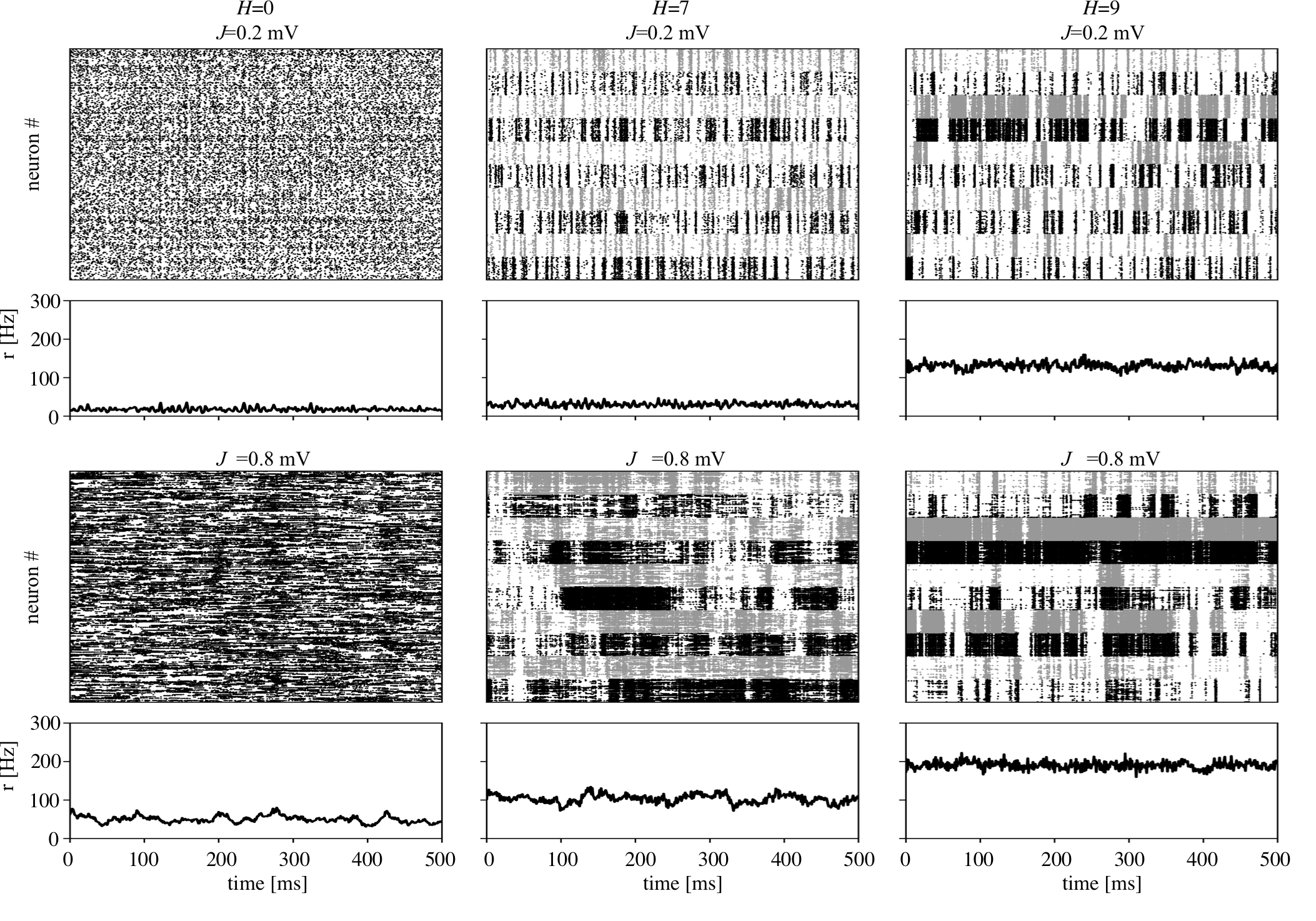}
	\caption{\textbf{Raster plot and activity plot of the network for selected values of $J$ and $H$.} For visibility, raster plots show spike times for a sample of only 2560 neurons but the activity plots refer to all neurons in the network. Each column corresponds to a hierarchical level (from left to right: $H=0$, $H=7$, $H=9$), and each row corresponds to a synaptic strength (upper row: $J=0.2$ mV; bottom row: $J=0.8$ mV). In the cases of modular networks ($H=7$ and $H=9$), spikes of neurons in the same module are indicated by the same color (black or gray), which alternate from one module to another to ease visualization. Although modules in the network with $H=9$ have smaller number of neurons than modules in the network with $H=7$, the same number of neurons per module was chosen for the cases of $H=7$ and $H=9$ to allow a comparison.}
	\label{Fig:rasters}
\end{figure*}

As described in Methods, for each hierarchical level $H$ (in the range from 0 to 9) we ran simulations of the network with coupling strength $J$ in the range [0.1, 0.15, \ldots, 1] (in millivolts) and $g=5$. In Fig.~\ref{Fig:rasters} we show the raster plots and corresponding firing rates for three $H$ values ($H=0$, which corresponds to an Erd\H{o}s-R{\'e}nyi graph; $H=7$; and $H=9$) and two $J$ values ($J=0.2$ mV and $J=0.8$ mV).    

The network with $H=0$ can have two types of asynchronous activity. In the case of week coupling (cf. $H=0$ and $J=0.2$~mV in Fig.~\ref{Fig:rasters}), neurons fire irregularly and no synchronous behavior is observed. In addition, the population firing rate is low (the average value of $r(t)$ for $J=0.2$~mV is $\nu = 17.6\pm5.6$ Hz, where the $\pm$ sign means standard deviation) and homogeneous. As the synaptic strength increases (cf. $H=0$ and $J=0.8$~mV in Fig.~\ref{Fig:rasters}), the activity changes to a more heterogeneous behavior where single neurons fire in bursts of high activity interspersed with short periods of low activity, and the network firing rate displays a less homogeneous behavior with some irregular fluctuations. The mean firing rate also increases ($\nu = 53.1\pm12.5$ Hz for $J=0.8$~mV). An evidence of the fluctuations that appear when $J$ is increased is the growth of the standard deviation of $r(t)$, which more than doubles when $J$ changes from $0.2$~mV to $0.8$~mV.

In the second and third columns of Fig.~\ref{Fig:rasters} we compare activity dynamics for hierarchical levels $H=7$ and $H=9$ and synaptic strengths $J=0.2$~mV and $J=0.8$~mV. For both hierarchical levels, heterogeneous spiking behavior and modularity effects appear already for low synaptic strength (cf. $J=0.2$~mV) and become more pronounced as $J$ increases (cf. $J=0.8$~mV). The population firing rate also is very sensitive to increases in both $J$ and $H$. For fixed $J$ the firing rate increases with $H$, and for fixed $H$ the firing rate increases with $J$. For quantitative comparison, the average population firing rate values are: (i) ($H=7$, $J=0.2$~mV): $\nu =30.2\pm7.7$ Hz; (ii) ($H=7$, $J=0.8$~mV): $\nu =102.9\pm15.4$ Hz; (iii) ($H=9$, $J=0.2$~mV): $\nu =129.3\pm12.1$ Hz; and (iv) ($H=9$, $J=0.8$~mV): $\nu =187.8\pm16.6$ Hz.     
In addition to that, as $H$ increases modules begin to act more individually as can be seen in the different spike patterns of each module (observe the horizontal bands in alternating gray and black colors for panels with $H=7$ and $9$). In the following, we will show that both high hierarchical level $H$ and high synaptic strength $J$ also increase information transmission in the network. 

\begin{figure*}[!htp]
	\centering
	\includegraphics[scale=0.35]{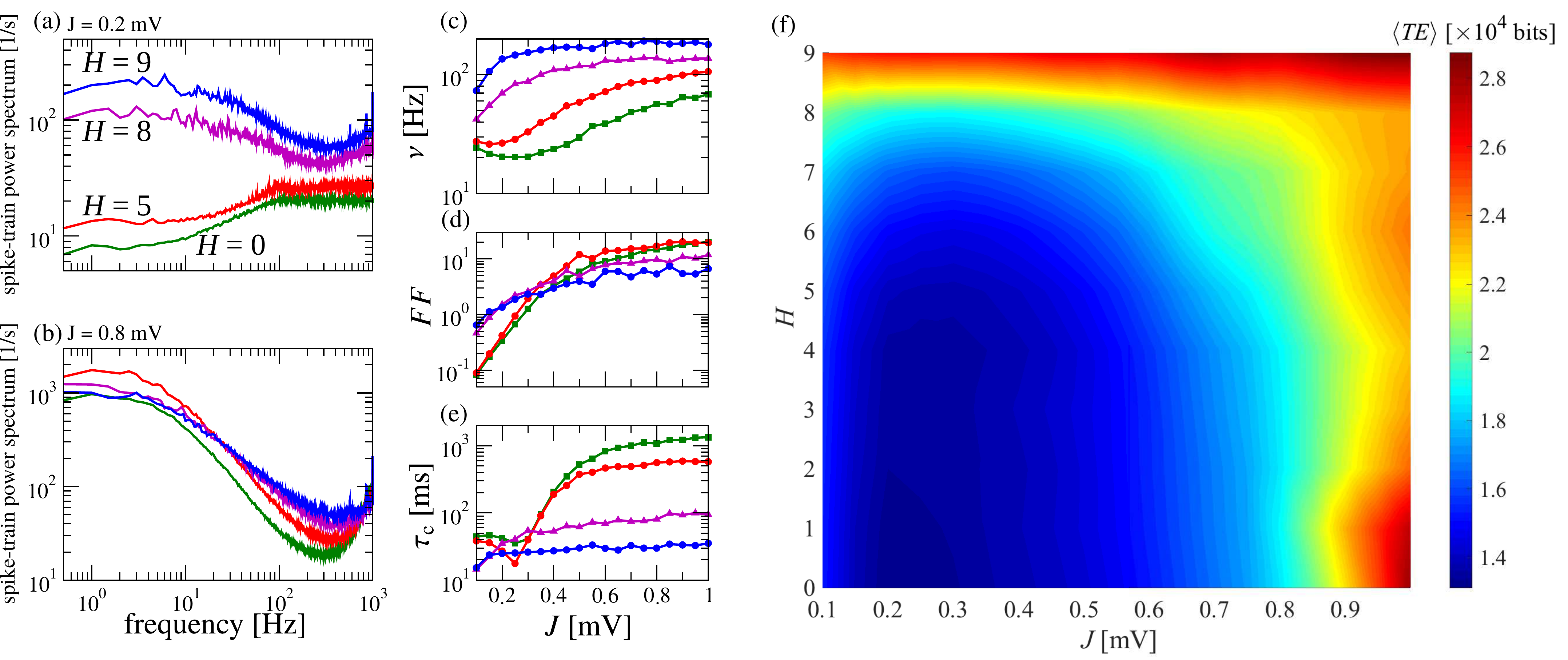}
	\caption{\textbf{Increases of $J$ and $H$ cause amplification of slow fluctuations and enhance information transfer.} (a) Spike-train power spectra computed for $J=0.2$~mV and different values of $H$ (indicated by different colors in the plot). (b) Same plot as in (a) but with $J=0.8$~mV. (c--e) Firing rate $\nu$, Fano factor $FF$, and correlation time $\tau_{\rm c}$ for different values of $J$ ($H$ values indicated by the same colors as in (a,b)). (f) Average transfer entropy (computed as in Eq.~\ref{Eq:TE_mean}) in a two-dimensional diagram where the abscissa represents synaptic strength $J$ and the ordinate represents hierarchical level $H$. Values of $\langle TE \rangle$ are indicated by the color bar to the right side.}
	\label{Fig:statistics}
\end{figure*}

In Figs.~\ref{Fig:statistics}(a--e) we present extended statistics that shed light on the effects of increasing $J$ and $H$. Analysis of the spike-train power spectra in Figs.~\ref{Fig:statistics}(a,b) shows that an increase of either $J$ or $H$ leads to a build-up of slow fluctuations in the network. However, the effect is more pronounced for an increase in $J$ than for an increase in $H$. For example, for fixed $H=0$ a change in $J$ from $0.2$~mV to $0.8$~mV produces increases in power at low frequencies of about 2 orders of magnitude, whereas for fixed $J=0.2$~mV a change in $H$ from $0$ to $9$ produces power increases at low-frequencies of about 1 order of magnitude. Overall, the spectral characteristics are similar to the ones of cortical neurons \cite{bair1994}. 

For low values of $H$, typically $H<7$, the mean network firing rate $\nu$ displays non-monotonic behavior as a function of $J$. It initially decreases towards a minimum and then increases as shown in Fig.~\ref{Fig:statistics}(c) (curves in green and red). The minimum marks the transition from the asynchronous homogeneous behavior to the asynchronous heterogeneous behavior (compare the raster plots in Fig.~\ref{Fig:rasters} for $H=0$.) For  $H \geq 7$ the minimum disappears and the curve of $\nu$ versus $J$ grows monotonically towards a saturation firing rate (purple and blue curves in Fig.~\ref{Fig:statistics}(c)). 

The Fano factor $FF$, on the other hand, grows with $J$ for all hierarchical levels $H$. What changes is the growth rate, which is much higher for low $H$ than for high $H$ (again, the transition point is around $H=7$). For low $H$, $FF$ starts at values well below 1 (indicating low spike variability) for low synaptic strengths and rises steeply to values about two orders of magnitude higher as the synaptic strength increases, indicating a rapid increment in spike variability (see green and red curves in Fig.~\ref{Fig:statistics}(d)). The $FF$ growth is not so pronounced when $H \geq 7$, with variations of one order of magnitude or less (purple and blue curves in Fig.~\ref{Fig:statistics}(d)). Interestingly, the asymptotic $FF$ value for large $J$ is lower for $H=9$ than for $H=8$, suggesting that there is a limiting level of modularity beyond which spike variability and heterogeneity do not grow.

The behavior of the correlation time $\tau_{\rm c}$ as a function of $J$ is similar to the one of the firing rate $\nu$. It decreases to a minimum and then increases with $J$ when $H<7$, and grows monotonically with $J$ for $H \geq 7$ (Fig.~\ref{Fig:statistics}(e)). Overall, the behavior of $\nu$, $FF$ and $\tau_{\rm c}$ reflect the amplification of slow fluctuations and increments of network firing rate and spike variability provoked by topological (introduction of modularity) and synaptic strength changes in the network, and are comparable with the behavior of these variables for random networks with fixed in-degrees reported elsewhere \cite{wieland2015,pena2018}.

In order to characterize information flow in the network, we show in Fig.~\ref{Fig:statistics}(f) the behavior of  $\langle TE\rangle$ in the parameter space spanned by $J$ and $H$ (each point corresponds to an average over 10 different initial conditions). For very low values of synaptic coupling ($J \lessapprox 0.2$), the effect of modularity on $\langle TE\rangle$ is not very significant until $H \gtrapprox 6$, as can be seen from the vertical arrangement of shaded stripes in the diagram. Then, for intermediate coupling strengths ($0.2 \lessapprox J \lessapprox 0.5$) the effect of modularity on $\langle TE\rangle$ becomes significant (stripes are predominantly horizontal), and, for strong coupling ($J \gtrapprox 0.5$), the effect is again reduced (stripes are vertically arranged again). The exception is
when the number of modules is very high ($H \geq 8$), in which case $\langle TE\rangle$ is insensitive to coupling strength. Regarding the behavior of $\langle TE\rangle$ with respect to changes in $J$ and $H$, in the region of the diagram where $\langle TE\rangle$ is more sensitive to $J$ (region with $H \leq 5$) $\langle TE\rangle$ decreases towards a minimum as $J$ grows from 0.1 to 0.3, and then increases toward high values as $J$ grows from 0.3 to 1. This behavior is similar to the one for $\tau_{\rm c}$ depicted in Fig.~\ref{Fig:statistics}(e). The maximum value of $\langle TE\rangle$ in this region occurs for strong coupling ($J = 1$) and either no modules ($H = 0$) or only two modules ($H = 1$). And in the region of the diagram where the effect of modularity is important ($H \geq 5$), $\langle TE\rangle$ tends to grow with $H$. The maximum value of $\langle TE\rangle$ is attained for the largest number of modules considered ($H = 9$), and this value is comparable to the maximum of $\langle TE\rangle$ in the region where $\langle TE\rangle$ is more sensitive to $J$.    

Results in this section show that both slow fluctuations and information transmission are largely enhanced when  $J$ and $H$ grow. We hypothesize that, as $J$ and $H$ increase modules start to act as single units. For example, in Fig.~\ref{Fig:rasters} the modules in networks with high $J$ and $H$ exhibit different individual behavior and can be identified visually. All modules display bursts of intense activity intercalated with periods of low activity, but each module has its own pattern of burst\slash quiescence alternations which does not coincide with the patterns of the others. This is suggestive that when both synaptic coupling and the number of modules are high, modules behave as independent functional units. In the next section we investigate this suggestion by studying the auto- and cross-correlations of the neuronal spike-trains.

\subsection{Effects of {\it J} and {\it H} on the autocorrelation and cross-correlation of single-neuron spike-trains}

In this section, we investigate the autocorrelation and cross-correlation of the spike-trains of single neurons in order to obtain a better understanding of the individual properties of neurons when slow fluctuations and information transmission are incremented due to increases in the synaptic coupling strength $J$ and\slash or the hierarchical level $H$. 

\begin{figure}[!htp]
	\centering
	\includegraphics[scale=0.55]{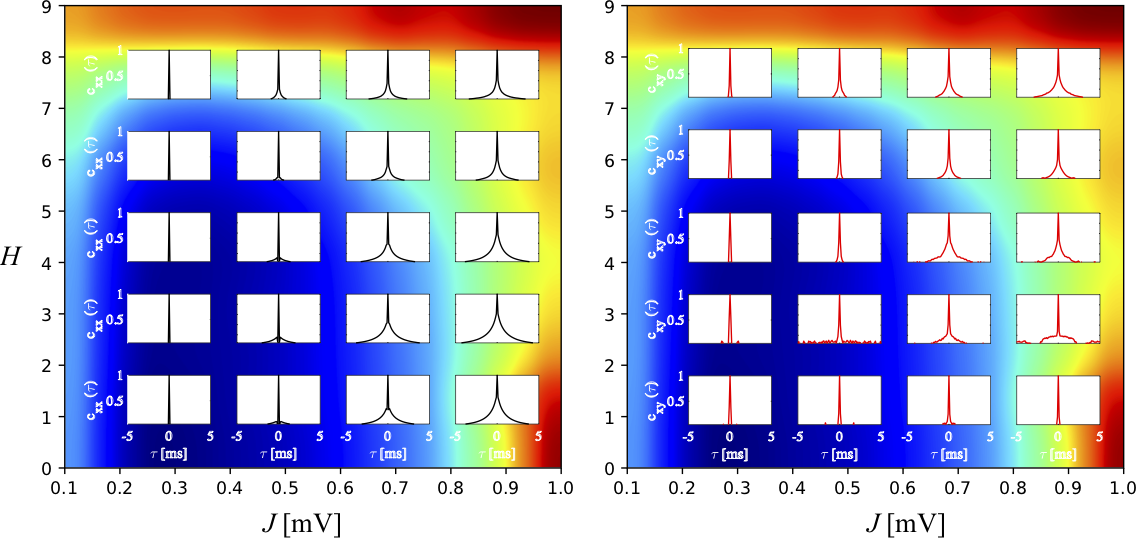}
	\caption{\textbf{Spike-train autocorrelation $c_{xx}(\tau)$ and cross-correlation $c_{xy}(\tau)$ for selected pairs of parameters ($H$,$J$)}. Left: $c_{xx}$. Right: $c_{xy}$. The selected pairs ($J$,$H$) correspond to all possible combinations taken from the sets $J=\{0.2,0.4,0.6,0.8\}$ and $H=\{0,2,4,6,8\}$. For better visualization, $c_{xx}$ and $c_{xy}$ for the pairs ($J$,$H$) are plotted over the plot of $\langle TE\rangle$ in the $J$-$H$ diagram.  The $c_{xx}$ is extracted from $K=10,000$ randomly chosen neurons and the $c_{xy}$ from  $K=10,000$ randomly chosen pairs of neurons.}
	\label{Fig:auto_cross}
\end{figure}

In Fig.~\ref{Fig:auto_cross} we show the autocorrelation $c_{xx}(\tau)$ and the cross-correlation $c_{xy}(\tau)$, as defined in Methods, for selected pairs of parameters ($J,H$) taken from the sets $J=\{0.2,0.4,0.6,0.8\}$ and $H=\{0,2,4,6,8\}$.  When the topology of the network is not modular (bottom row of Fig.~\ref{Fig:auto_cross}), the increase in the synaptic coupling $J$ produces an increase in the spike-train autocorrelation but has almost no effect on the spike-train cross-correlation. This reflects the effect of $J$ in enhancing slow fluctuations while keeping the network activity asynchronous as observed before (cf. the first column of the raster plots in Fig~\ref{Fig:rasters} and the curves for $H = 0$ (green curves) in Figs.~\ref{Fig:statistics}(a--e)). In other words, in a non-modular network, when the synaptic coupling increases the spikes of an individual neuron tend to become more correlated over short times but behave independently of the spikes of other neurons. 

In contrast to this situation, when the number of modules is high (upper rows of Fig.~\ref{Fig:auto_cross}) the increment in $J$ affects both the spike-train autocorrelation and cross-correlation. The cross-correlation over a short-time increases when the synaptic coupling is strong, indicating a weak but non-negligible degree of functional coupling between neurons. In addition, the autocorrelation also increases with $J$ but now this increase is less pronounced than when $H=0$. 

The different behaviors of the spike-train auto- and cross-correlations upon increment in $J$ between networks with non-modular and modular topologies hints that a more complex activity pattern emerges at the population level when hierarchical modularity is introduced in the network, which was not present when $H=0$. Moreover, the microscopic $\langle TE\rangle$ measured used in the previous section was not able to capture this difference: in the diagram of Fig.~\ref{Fig:statistics}(f) the regions defined by ($H=0$, $J \geq 0.9$) and ($H=0.9$, $J \geq 0.9$) have approximately the same values of $\langle TE\rangle$.    
The above results suggest that the introduction of a hierarchical modular topology produces some form of population communication (reflected in the increase of spike-train cross-correlation) that was not present in the network with non-modular topology. Since the $\langle TE\rangle$ measure was not sensitive to this finding, we will use the macroscopic $TE$ ($\langle TE^{(H)} \rangle$) introduced in Methods to test whether it can be helpful in this case. This is the subject of the next section.

Why does the spike-train cross-correlation increases with the hierarchical level? In order to understand this, below we derive equations to investigate how the internal (i.e. intramodular) and external (i.e. intermodular) communication is affected by the hierarchical level $H$. We focus on the average number of connections as they are rewired at any new increment in $H$. In the calculations below we will not make any distinction between excitatory/inhibitory connections, thus keeping everything in general terms.

Let us start with the network where $H=0$. For large $N$, the expected number of connections to a neuron which come from inside the single module is $n_{\rm in}^{(H=0)} = N \epsilon$, where the superscript indicates the hierarchical level $H=0$.

Now, when $H=1$ the rewiring algorithm tells that one should divide the network and rewire its connections, which means that the expected number of connections to a neuron from the same module where it is located is half of the previous value plus the expected number of connections to the other module that are cut and rewired back to the neuron (we will assume, for simplicity, that the rewiring probability is $R$ for all connections): 
\begin{equation}
\label{Eq:n1}
n_{\rm in}^{(H=1)} = \frac{n_{\rm in}^{(H=0)}}{2} + \frac{n_{\rm in}^{(H=0)}}{2}\times R.
\end{equation}

Eq.~\ref{Eq:n1} gives the average number of connections to a neuron that come from inside the same module. In a similar way, the average number of connections that come from outside the module to the neuron is given by 
\begin{equation}
\label{Eq:no1}
n_{\rm out}^{(H=1)} = n_{\rm in}^{(H=0)} - n_{\rm in}^{(H=1)} = N \epsilon - n_{\rm in}^{(H=1)}.
\end{equation}

Note that we can re-write Eq.~\ref{Eq:no1} for any hierarchical level $H>0$ because the expected number of connections from outside a module will always be the expected number of connections at $H=0$ minus the expected number of connections from inside the module after rewiring:
\begin{equation}
\label{Eq:nom}
n_{\rm out}^{(H)} = N \epsilon - n_{\rm in}^{(H)}.
\end{equation}

For the hierarchical level $H=2$, we follow the same procedure used to derive equation~\ref{Eq:n1} and obtain the expression for $n_{\rm in}^{(H=2)}$, but now considering that the connections from outside the module when $H=1$ are also rewired:
\begin{align}
	\label{Eq:n2}
	n_{\rm in}^{(H=2)} & = \frac{n_{\rm in}^{(H=1)}}{2} + \frac{n_{\rm in}^{(H=1)}}{2}\times R + n_{\rm out}^{(H=1)}\times R \nonumber \\
	& = \frac{n_{\rm in}^{(H=1)}}{2}(1-R) + N \epsilon \times R.
\end{align}

For hierarchical levels $H>1$, we recursively apply the above equations and obtain the expression 
\begin{align}
	\label{Eq:h_plus_1}
	n_{\rm in}^{(H+1)} & = \frac{N \epsilon}{2} \left[\left(\frac{1-R}{2}\right)^H + 2R\sum_{k=0}^H\left(\frac{1-R}{2}\right)^k \right].
\end{align}

In summary, Eq.~\ref{Eq:h_plus_1} gives the expected number of connections to a neuron that comes from its own module at the hierarchical level $H>1$, and Eq.~\ref{Eq:nom} gives the expected number of connections to a neuron that comes from outside its module for any $H>0$. 

It is interesting to note that the rewiring procedure is limited with respect to $n_{\rm in}$, so that $\lim_{H\rightarrow\infty} n_{\rm in} = \frac{2NR\epsilon}{R+1}$. This means that while increasing $H$, the average number of connections to a neuron that come from inside the same module reaches a fixed value, no matter how small is the module. This fact is important because it shows that the average density of connections ($\epsilon_{\rm in} = (2^{H} \times n_{\rm in} ) / N$) in a module increases dramatically when such a limit is achieved since the number of neurons within a module decreases as $H$ increases. Concomitantly, $n_{\rm out}$ is also limited since it is directly related to $n_{\rm in}$.

\begin{figure}[!htp]
	\centering
	\includegraphics[scale=0.7]{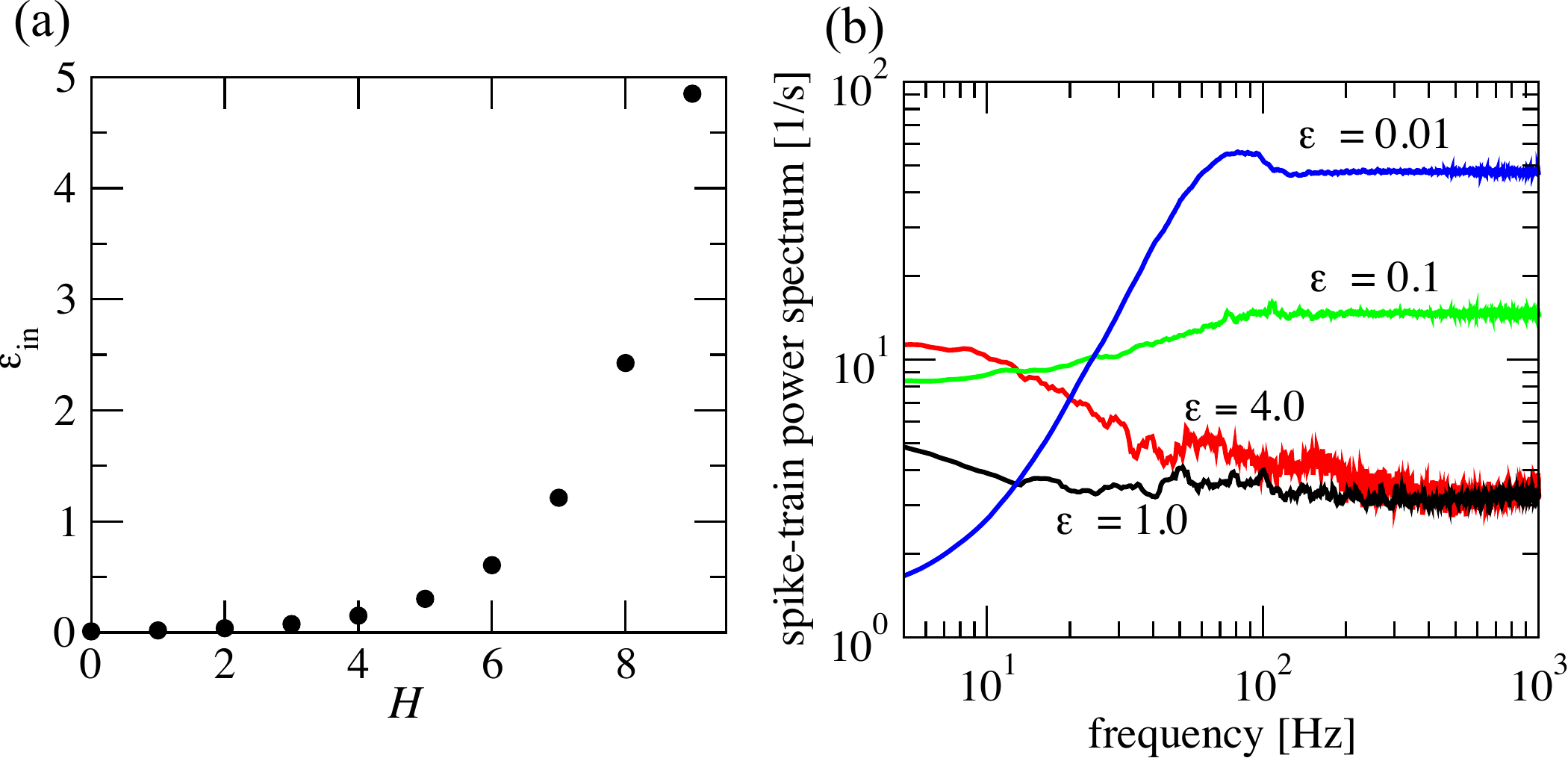}
	\caption{\textbf{Relation of connectivity and slow fluctuations.} (a) Values of connectivity inside a module ($\epsilon_{\rm in}$) as $H$ increases (cf. Eqs.~\ref{Eq:n1}~--~\ref{Eq:h_plus_1}). (b) Spike-train power spectra extracted for a small network with $N=2^{14}$ and $H=0$ for different values of $\epsilon$.}
	\label{Fig:conns_epsilon}
\end{figure}

The set of Eqs.~\ref{Eq:n1}~--~\ref{Eq:h_plus_1} can elucidate why cross-correlations increase in a module as $H$ increases. In Fig.~\ref{Fig:conns_epsilon}(a) we show how the value of $\epsilon_{\rm in}$ changes as a function of the hierarchical level $H$. One can see that connections within a module grow exponentially with $H$. As $\epsilon_{\rm in}$ exponentially increases, a higher degree of synchronous activity in the network is expected, and thus correspondingly higher values of spike-train cross-correlations are also expected. In fact, it is expected that a random rewiring of connections, which is equal in nature to random occurrences of events in a Poisson process, would lead to a exponential growth of spike-train cross-correlations.

To check how slow fluctuations build up with increasing connectivity within a module, we simulated a network with $N=2^{14}$ neurons and $H=0$ (representing a single module) with varying values of $\epsilon$. The spike train power spectra of the network for the different values of $\epsilon$ are shown in Fig.~\ref{Fig:conns_epsilon}(b). One can see that slow fluctuations start to build up as $\epsilon$ increases (note the initial values on the left hand side of the plots). 

Results in this section show how the single-neuron behavior is affected by increases of $J$ and $H$. Some phenomena, like the enhancement of information transfer and the build up of slow-fluctuations, emerge and display similar properties when either $J$ and $H$ are large. However, other measures like the spike-train autocorrelation and cross-correlation behave in  different ways when either $J$ or $H$ increase. In particular, the results suggest that information flow at the population level is more robust in the presence of a hierarchical and modular network. To understand better how information flow at the population level is affected when the hierarchical level is increased, in the next section we study the effect of increasing $J$ and $H$ on the macroscopic $TE$ introduced in Methods.

\subsection{Information flow at the population level}

In this section we focus on how information flows at the macroscopic scale of modules in the network. The algorithm used to build hierarchical modular topologies allows to gradually observe how different measures increase or decrease with the parameter $H$. We have already shown that $H$ and $J$ affect differently the spike-train auto- and cross-correlations, and in this section we are interested on how information flow measured at the modular level behaves as $J$ and $H$ vary. Is the behavior different or similar to the one seen for information flow at the single-neuron level? 

First, we recall Fig.~\ref{Fig:statistics}(f), where it can be observed that increasing $H$ causes an enhancement in information flow at the microscopic level ($\langle TE\rangle$). This can be interpreted as an increase in the ``usefulness'' of the knowledge of the spike train of a give neuron in predicting the future behavior of the spike train of a different neuron. Here, considering the hypothesis that communication can take place not only at the level of the single units of the network ("microscopic" level) but also at the level of the modules in which the network is organized ("macroscopic" level), we will evaluate information flow among modules using the measure $\langle TE^{(H)} \rangle$ introduced in the Methods section. 

\begin{figure}[!htp]
	\centering
	\includegraphics[scale=0.3]{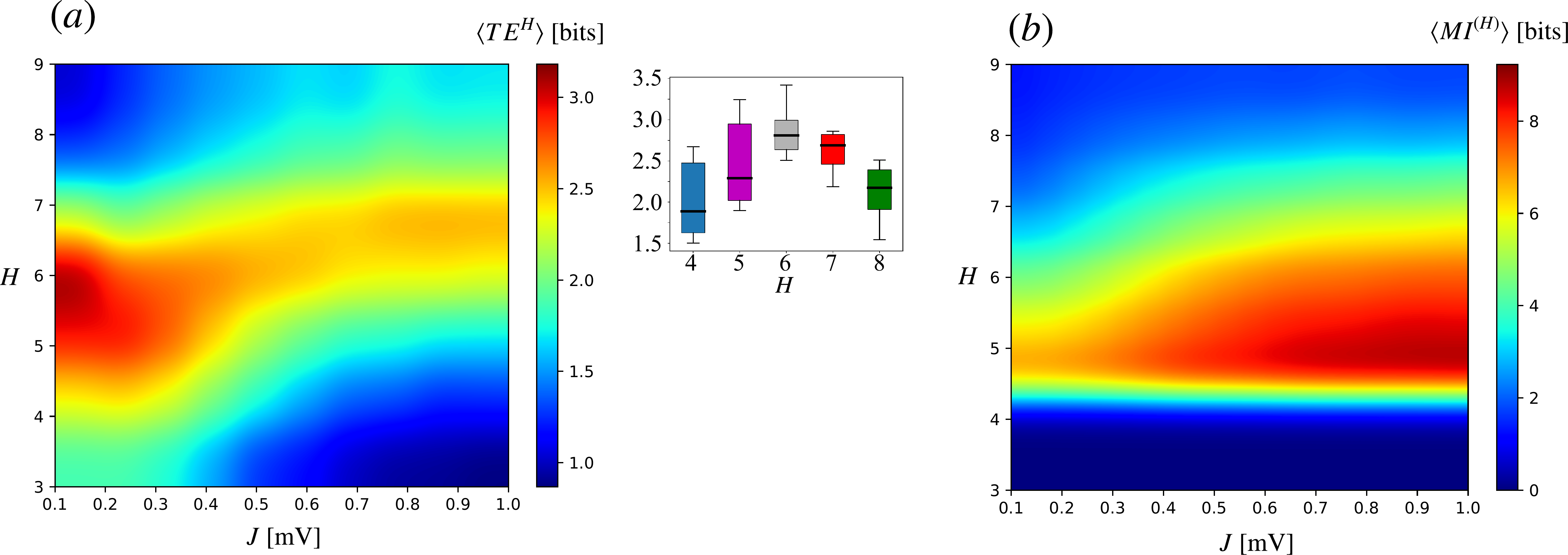}
	\caption{\textbf{Transfer entropy and mutual information among modules.} (a) Transfer entropy evaluated among modules  $\langle TE^{(H)} \rangle$ in the two-dimensional diagram where the ordinate represents the hierarchical level $H$ and the abscissa represents the synaptic strength $J$. Inset: boxplots of $\langle TE^{(H)} \rangle$ for fixed values of $H$. (b) Mutual information among modules $\langle MI^{(H)} \rangle$ in the same $J$-$H$ diagram.}
	\label{Fig:te_atv}
\end{figure}

In Fig.~\ref{Fig:te_atv}(a) we can observe that the communication among modules is indeed very different from the one among neurons shown in Fig.~\ref{Fig:statistics}(f). The most compelling difference is the existence of an intermediate range of $H$ values (around $H=6$) at which $\langle TE\rangle$ is maximal. Also, above and below this range there are two contrasting behaviors: for low $H$ ($H \leq 4$), $\langle TE\rangle$ monotonically decays with $J$ as $J$ increases; for high $H$ ($H \geq 7$) this behavior is somewhat mirror-inverted and $\langle TE\rangle$ monotonically increases with $J$. 

The boxplots at the inset of Fig.~\ref{Fig:te_atv}(a), which display the distributions of $\langle TE^{(H)}\rangle$ for different $H$ values and the entire range of $J$ values, show that $H=6$ has the highest mean and the lowest variance of $\langle TE^{(H)}\rangle$. This clearly shows that $H=6$ is an optimized point for information transmission among modules.  

The results in Fig.~\ref{Fig:te_atv}(a) indicate that a form of modular communication takes place in the hierarchical modular networks. There is an "optimal" level of hierarchical modular organization (neither the lowest nor the highest level) at which the macroscopic $TE$ is maximal. Moreover, at this "optimal" $H$ level the macroscopic $TE$ is relatively insensitive to changes in the synaptic strength $J$. Only when $H$ is above or below the optimal value the communication at modular level is significantly influenced by the synaptic strength $J$. 

Results of the previous two sections suggest that as $H$ increases the modules start to behave as individual functional units. To test this hypothesis we computed the mutual information among modules, $\langle MI^{(H)} \rangle$. This metric can be interpreted as a measure of statistical dependence among the considered elements \cite{deabril2018}. In Fig.~\ref{Fig:te_atv}(b) (neglecting the behavior for $H \leq 4$) one can see that as $H$ increases $\langle MI^{(H)} \rangle$ decreases indicating that the modules act more independently as the hierarchical modular level increases. Interestingly, Fig.~\ref{Fig:te_atv}(b) also shows that for intermediate $H$ values ($5 \leq H \leq 7$) the synaptic strength $J$ plays a role on the statistical dependence among modules. Within this intermediate range of $H$ values, $\langle MI^{(H)} \rangle$ increases with $J$ indicating that the modules become less statistically independent as the synaptic strength increases. Since the microscopic parameter $J$ is associated with the emergence of slow fluctuations in the network activity, this points to a link between slow activity fluctuations and statistical dependency among modules.  

\section{Discussion}

An important problem in computational neuroscience is the investigation of different dynamics displayed by networks of spiking neurons \cite{brunel2000,RenDel10,wang2011,pena2018b} 
and in particular the ones that enhance information processing such as dynamics with slow fluctuations \cite{LitDoi12,ostojic2014,wieland2015}. Region-to-region communication characteristics and how they interact with the topological features of the network are also of great interest because they shed light on the relationship between topology and dynamics \cite{SpoChi04,reijneveld2007}. Here, we addressed this problem by investigating networks with hierarchical modular topology, which display generic features of cortical networks \cite{mountcastle1997,kaiser2010,TomPen14}, and how the topological structure affects information flux. 

We have constructed large networks of spiking neurons with variable levels of (i) hierarchy and modularity, and (ii) synaptic strength. By extracting information-theoretic measures (microscopic and macroscopic $TE$ and $MI$), we were able to observe that both information propagation and slow activity fluctuations can be optimized by combining (i) and (ii). Our goal was to analyze how the interplay of intrinsic neuronal parameters and topological features influences activity propagation and how this is related to different spatial scales (the "microscopic" scale of single neurons and the "macroscopic" scale of neuronal modules).

More specifically, we started with a comparison of spiking activity characteristics between networks with Erd\H{o}s-R{\'e}nyi and hierarchical modular topologies. The activities of the networks with the two topologies were characterized in terms of their variation with the synaptic strength $J$. Since the relative inhibitory synaptic strength $g$ is fixed in 5, previous works have already shown that the activity displayed by these networks is of the type known as "asynchronous irregular" (AI) \cite{ostojic2014,wieland2015,pena2018}. Indeed, we have observed AI-like activity in our networks. In networks with AI activity, neurons fire without correlation and the increase of $J$ to high values creates a second type of AI activity, called "heterogeneous" AI \cite{ostojic2014}, which is characterized by the emergence of slow fluctuations \cite{wieland2015,pena2018}. The heterogeneous AI regime has bursts of spikes intercalated with periods of silence. We observed this pattern again in our study but for high values of the hierarchical level $H$ the heterogeneous behavior appears even at low $J$. Moreover, when $H$ is high the different modules display heterogeneous spiking patterns, i.e. they behave as units independent from each other.

Then, we moved on to a study of information transmission in the hierarchical modular networks as a function of the topological parameter $H$ and the microscopic synaptic strength parameter $J$. To investigate possible different ways of communication in the network, namely at the microscopic level of neurons and at the macroscopic level of modules, we used two different measures of $TE$: $\langle TE\rangle$ and $\langle TE^{(H)}\rangle$. The microscopic measure $\langle TE\rangle$ is based on the neuronal spike trains, and the macroscopic measure $\langle TE^{(H)}\rangle$ is based on the average firing rates (activities) of the modules.  

Let us call the type of communication at microscopic level $C_{\mbox{micro}}$ and the type of communication at macroscopic level $C_{\mbox{macro}}$. Then, when exploring $C_{\mbox{micro}}$ and $C_{\mbox{macro}}$ we had two possibilities: (i) $TE$ in $C_{\mbox{macro}}$ is predictable from the measurement of $TE$ in $C_{\mbox{micro}}$ (and vice-versa); or (ii) communication at these two scales is completely different. If possibility (i) were true, we would expect that the two measures, $\langle TE\rangle$ and $\langle TE^{(H)}\rangle$, would display similar properties when observed in the $J$-$H$ diagram. In such case, communication in the network would be independent of the two scales and bridging between $C_{\mbox{micro}}$ and $C_{\mbox{macro}}$ would be directly possible. On the other hand, if possibility (ii) were true knowledge of either $\langle TE\rangle$ or $\langle TE^{(H)}\rangle$ could not be used to explain the other measure because they would be capturing different things.    

Our study has shown that possibility (ii) is true, i.e. $C_{\mbox{micro}}$ and $C_{\mbox{macro}}$ are different. The behavior of $\langle TE\rangle$ in the $J$-$H$ diagram shows that there are two regions where $C_{\mbox{micro}}$ is maximal: the line on top of the diagram where $H=9$ (independent of $J$), and the bottom right-hand corner where $H \leq 1$ and $J \approx 1$. The $J$-$H$ diagram for $\langle TE^{(H)}\rangle$ shows an opposite situation: $C_{\mbox{macro}}$ is maximal along the line given by $H=6$ and is very low at the regions where $C_{\mbox{micro}}$ is maximal. The main finding of our study is that there is an intermediate value of hierarchical level (within the range of $H$ values considered) for which $C_{\mbox{macro}}$ is maximal. This "optimal" type of behavior was not found when we studied $C_{\mbox{micro}}$.        

As an attempt to explain the observed behavior of $C_{\mbox{micro}}$ and $C_{\mbox{macro}}$, we investigated two other types of measures. In the case of $C_{\mbox{micro}}$, we used the spike-train auto- and cross-correlations. In the case of $C_{\mbox{macro}}$, since our hypothesis was that the observed behavior was due to the emergence of independent modules, we used the mutual information among modules, $\langle MI^{(H)} \rangle$.      

As noted above, in the $J$-$H$ diagram for $\langle TE\rangle$ there are two regions where $\langle TE\rangle$ is maximal: the upper right-hand corner where both $H$ and $J$ are highest and the lower right-hand corner where $H=0$ and $J=1$. The observation of $\langle TE\rangle$ alone is not enough to reveal the mechanisms underlying these seemingly similar behaviors. The use of the spike-train auto- and cross-correlations helps in this disambiguation. The high $\langle TE\rangle$ for a non-modular network with high $J$ is due to the increase in the spike-train autocorrelation with the increase of $J$, while the high $\langle TE\rangle$ for a network with high $J$ and many modules is due to the increase in the spike-train cross-correlation with the increase of $H$.   

Interpreting $\langle MI^{(H)} \rangle$ as a measure of independence among modules (high $\langle MI^{(H)} \rangle$ meaning higher relative dependence, and low $\langle MI^{(H)} \rangle$ meaning lower relative independence), our results (cf. Fig.~\ref{Fig:te_atv}(b)) show that modules become relatively more independent as $H$ grows (neglecting situations with $H \leq 4$). The situation with highest level of modular independence is the one with highest $H$ ($H=9$) and the situation with lowest level of modular independence is the one with lowest $H$ ($H=5$). Combining this result with the results shown in the diagram for $\langle TE^{(H)}\rangle$ in Fig.~\ref{Fig:te_atv}(a), one sees that the scenario with maximum $C_{\mbox{macro}}$ occurs in a situation where modules are neither too independent nor too dependent from each other. If all modules were completely independent they would act as autonomous units and $\langle TE^{(H)}\rangle$ would be near zero; if the modules were very interdependent, they would act more or less as a single unit and $\langle TE^{(H)}\rangle$ also would be low (knowledge of the activity of a single module would be enough to infer the activities of all the other modules). Therefore, the optimal situation for information transfer among modules as measured by $\langle TE^{(H)}\rangle$ is the situation in which modules are in an intermediate position between total autonomy and total interdependence. This corresponds to the case with $H=6$.  

The optimal value $H=6$ does not mean that there is something special about the number 6. Our study only shows that the modular $TE$ is maximized at an intermediate value in the range of $H$ values used, which in our case was $[0, 9]$ because of the number $N$ of neurons chosen. We predict that a similar study with twice as many neurons, which would allow $H$ values close to 20, would result in an optimal $H$ value higher than 6.

Previous studies have concentrated either on other features that are enhanced by topological characteristics or on different types of activity regimes. For instance, it has been shown that hierarchical modular networks are advantageous for long-lived self-sustained activity \cite{TomPen14,TomPen16} and can present critical behavior \cite{wang2011} that is related to optimal dynamic range \cite{KinCop06}. Complementary to that, it has been shown that augmentation of the synaptic strength generates different versions of the standard AI activity which may favor information processing \cite{ostojic2014}. In our work, we have shown that hierarchical modularity also affects information transmission. In particular, our results suggest that there may be a transition point in the level of hierarchical modular organization which endows the network with high level of macroscopic communication independently of the synaptic strength. 

We have observed that slow activity fluctuations increase with both the hierarchical modular level $H$ and the synaptic strength $J$. However, the spike-train cross-correlation variation is more sensitive to $J$ than to $H$. Recent studies have investigated the influence of correlations in neuronal activity over information transmission \cite{GalFou06,MorRen08,barreiro2018}. Here, the used transfer entropy measure undoubtedly showed an increase in the information propagation at the single-neuron level at high hierarchical modular levels, which we showed to be related to the increase of the spike-train cross-correlation through the rewiring process. 

As one of the objectives of our work was to understand the benefits of a hierarchical modular structure for information transmission, we compared the microscopic $TE$, based on spike trains of pairs of neurons, with the macroscopic $TE$, based on firing rates of pairs of modules. Our results suggest that networks with hierarchical modular structure may be optimized for communication at the macroscopic level, i.e. at the level of modules instead of single neurons. A speculative interpretation of this is that signals produced at the level of modules (firing rates) are more robust and less prone to deleterious noise effects than signals produced at the level of single neurons (isolated spikes). 

In addition to that, our result that modules start to act more individually as the hierarchical modular level increases can be interpreted in line with suggestions made elsewhere that activity in modular networks provides functional segregation and integration \cite{sporns2000,wang2011}, which is certainly an advantage in terms of memory storage.

One final point concerning the difference between communication at micro and macro scales is worth mentioning. For communication at the level of spike-trains the information flow always increases with $J$, which would imply a high metabolic cost for synaptic communication \cite{vincent2003,harris2012}. On the other hand, for communication at the level of modular firing rates when the network is close to the optimal hierarchical level the variance of information flux is at a minimum, independently of the value of $J$. This suggests that the hierarchical modular structure may optimize the macroscopic information flow at a lower metabolic cost.

Overall, we believe that our work captures with a simple model novel important properties of communication and information processing in networks of spiking neurons. We provided new understanding on how topology may be connected to network dynamics (i.e. slow fluctuations) and information propagation. Our results and techniques can be applied to future research focused on how cortical networks optimize information processing and propagation.

\vspace{6pt} 

\funding{This paper was developed within the scope of the IRTG 1740 / TRP 2015/50122-0, funded by DFG / FAPESP. This work was partially supported by the Research, Innovation and Dissemination Center for Neuromathematics (FAPESP grant 2013/07699-0). RFOP is supported by a FAPESP PhD scholarship (grant 2013/25667-8), VL is supported by a CAPES Ph.D. scholarship. VL was partially supported by a FAPESP MSc scholarship (grant 2017/05874-0) at early stages of this work, ROS is supported by a FAPESP PhD scholarship (grant 2017/07688-9) and ACR is partially supported by a CNPq fellowship (grant 306251/2014-0). This study was financed in part by the Coordena\c{c}\~ao de Aperfei\c{c}oamento de Pessoal de N\'ivel Superior - Brasil (CAPES) - Finance Code 001.}

\conflictsofinterest{The authors declare no conflict of interest.} 

\appendixtitles{no} 
\appendixsections{multiple} 



\reftitle{References}





\end{document}